\documentclass[aps,prl,twocolumn,showpacs,superscriptaddress,groupedaddress]{revtex4-1}

\usepackage{graphicx}
\usepackage{dcolumn}
\usepackage{amssymb}
\usepackage{amsfonts}
\usepackage{amsmath}
\usepackage{bm}
\usepackage{color}
\usepackage{hyperref}
\usepackage[english]{babel}
\usepackage{version}
\usepackage{ulem}

\begin{document}

\title{Nucleation of ergodicity by a single mobile impurity in supercooled insulators}

\author{Ulrich Krause$^{1}$}
\author{Th\'eo Pellegrin$^{2}$}
\author{Piet W. Brouwer$^{1}$}
\author{Dmitry A. Abanin$^{2,3}$}
\author{Michele Filippone$^{2}$}

\affiliation{$^1$Dahlem Center for Complex Quantum Systems and Institut f\"ur Theoretische Physik, Freie Universit\"at Berlin, Arnimallee 14, 14195 Berlin, Germany}

\affiliation{$^2$Department of Quantum Matter Physics and $^3$Department of Theoretical Physics, University of Geneva, 24 Quai Ernest-Ansermet, CH-1211 Geneva, Switzerland}


\begin{abstract}

We consider a disordered Hubbard model, and show that, at sufficiently weak disorder, a single spin-down mobile impurity can thermalize an extensive initially localized system of spin-up particles. Thermalization is enabled by resonant processes which involve correlated hops of the impurity and localized particles. This effect indicates that certain localized insulators behave as ``supercooled" systems, with mobile impurities acting as ergodic seeds. We provide analytical estimates, supported by numerical exact diagonalization (ED), showing how the critical disorder strength for such mechanism depends on the particle density of the localized system. In the $U\rightarrow\infty$ limit, doublons are stable excitations, and they can thermalize mesoscopic systems by a similar mechanism. The emergence of an additional conservation law leads to an eventual localization of doublons. Our predictions apply to fermionic and bosonic systems and are readily accessible in ongoing experiments simulating synthetic quantum lattices with tunable disorder.

\end{abstract}


\maketitle


{\it Introduction --} Relaxation of a many-body system towards thermal equilibrium, driven by the interaction between its elementary constituents, is the cornerstone of statistical physics. Classically, thermalization is explained by the ergodic hypothesis, stating that isolated many-body systems forget  their initial conditions, exploring all possible configurations allowed by global conservation laws, such as energy conservation. The equivalent of the ergodic hypothesis in the quantum realm is the eigenstate thermalization hypothesis (ETH)~\cite{deutsch_quantum_1991,srednicki_chaos_1994,srednicki_approach_1999,polkovnikov_colloquium:_2011}.

It is of particular interest to find quantum systems avoiding thermalization. A generic mechanism  to violate ETH is provided by many-body localization (MBL)~\cite{basko_metalinsulator_2006,gornyi_interacting_2005,oganesyan_localization_2007,AbaninRMP}, which can be viewed as a generalization of the celebrated phenomenon of Anderson localization (AL)~\cite{anderson_absence_1958,abrahams_50_2010} to interacting systems, such as disordered Hubbard-type models studied experimentally with cold atoms~\cite{schreiber_observation_2015,choi_exploring_2016,lukin_probing_2019}. In MBL systems, the breakdown of thermalization stems from the emergence of local integrals of motion (LIOMs)~\cite{serbyn_local_2013,huse_phenomenology_2014,imbrie_many-body_2016}. They underlie surprising dynamical properties of MBL, which set it apart from AL, such as slow entanglement growth~\cite{znidaric_many-body_2008,bardarson_unbounded_2012,serbyn_universal_2013} and relaxation without thermalization~\cite{serbyn_quantum_2014}. 

In stark contrast to AL, which exists in dimensions $d=1,2$ at {\it any} disorder, MBL has been firmly established only at strong disorder in $d=1$. It is an open question how the transition from MBL to the ergodic phase occurs when disorder strength is reduced. Recent theories~\cite{vosk_theory_2015,potter_universal_2015,goremykina_analytically_2019} argued that this transition is driven by the formation of rare thermal ``bubbles". 
These theories build on a set of phenomenological assumptions regarding the interplay of bubbles and nearby, initially localized regions. It was shown that the bubbles can grow by including nearby localized degrees of freedom~\cite{de_roeck_stability_2017,luitz_how_2017,Eisert_avalanche,crowley_avalanche_2019}, which may destabilize certain localized systems, albeit at times that scale exponentially with the system size. However, these effects have not yet been observed in experiments with  quantum simulators, which can access local, time-dependent observables in localized systems~\cite{schreiber_observation_2015,choi_exploring_2016,lukin_probing_2019,chiaro_growth_2019,roushan_spectroscopic_2017}. It is therefore crucial to identify simple, realizable mechanisms for the generation and spreading  of thermal bubbles through an (initially) non-ergodic system.

In this study, we consider a paradigmatic correlated system -- disordered Hubbard model -- and uncover a novel  effect of interactions on localization. We consider a single spin-down impurity immersed in a spin-polarized background. We show that, similarly to seed crystals in supercooled liquids, at sufficiently weak disorder, a single impurity can act as ``ergodic nucleus'' that thermalizes the entire, initially localized system. 
For moderate on-site interaction strength $U$ of the order of the hopping $t$, mobile impurities take advantage of a finite density $\rho$ of single particles in adjacent localized states to propagate, destroying localization below a critical disorder strength $W_{\rm C}(\rho)$. In the infinite interaction limit, $U\rightarrow\infty$, a single doublon, which is a composite excitation made of two particles (two identical bosons or two fermions with opposite spin), stabilized by strong interactions~\cite{rosch_metastable_2008,strohmaier_observation_2010,macdonald_$fractu$_1988,hofmann_doublon_2012,wurtz_variational_2020}, can thermalize, via a similar mechanism, mesoscopic systems much larger than the single-particle localization length. The doublon is eventually localized due to the emergence of an additional conservation law.  

The mechanism proposed here applies to fermionic and bosonic systems, and differs qualitatively from that of Ref.~\cite{de_roeck_stability_2017}, which considered a static, well-thermalized region coupled to a localized system. In particular, we find delocalization time scaling as a power law in the system size, as opposed to an exponential scaling of Ref~\cite{de_roeck_stability_2017}. Furthemore, the nucleation mechanism can be probed by tracking the position of the impurity in out-of-equilibrium settings, which  could be readily realized in ongoing experiments on disordered optical lattices~\cite{schreiber_observation_2015,choi_exploring_2016,lukin_probing_2019,chiaro_growth_2019,roushan_spectroscopic_2017}. 



\begin{figure*}[t]
\includegraphics[width=.95\textwidth]{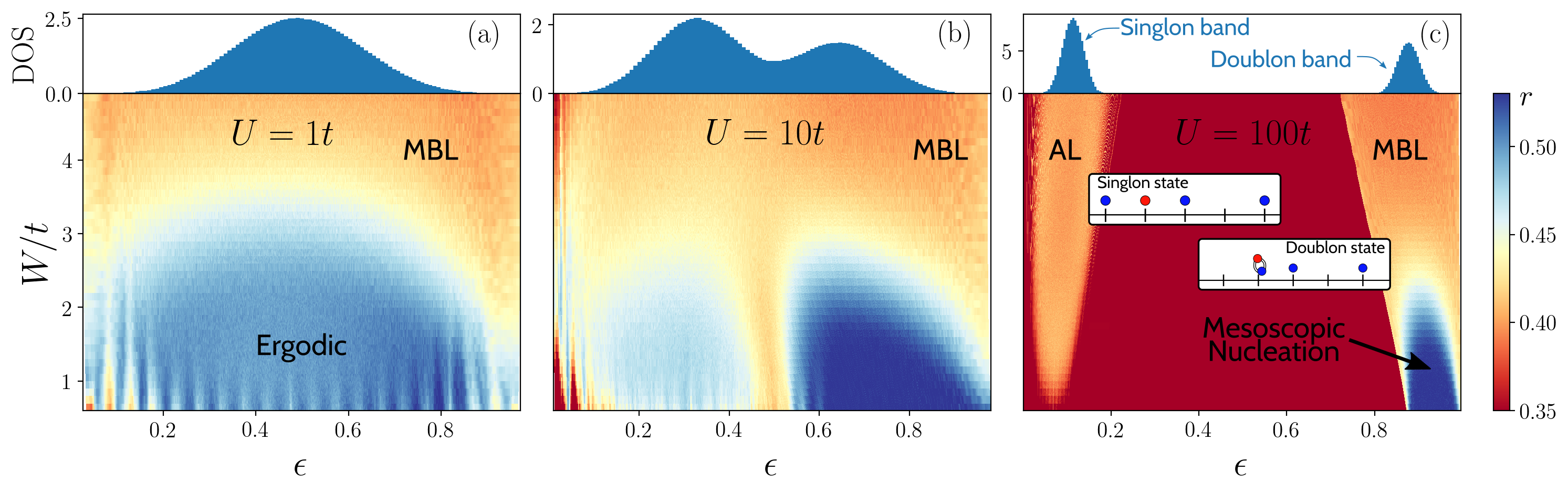}
\caption{Spectral-statistics parameter $r$ (bottom panels) for a system with a Hamiltonian (\ref{eq:fh}) with one spin-down impurity, shown as a function of disorder strength $W/t$ and renormalized energy  $\varepsilon=(E-E_{\rm min})/(E_{\rm max}-E_{\rm min})$. Panel (a) illustrates the case of moderate interaction $U=t$, while (b,c) show large-$U$ values, when a doublon can form. The upper panels show the many-body density of states (DOS) at $W=1.6t$, which exhibits a clear separation between sectors with and without doublon  at large $U$. At all values of $U$, a transition of Wigner-Dyson statistics (blue regions) at moderate disorder to Poisson statistics (yellow regions) is observed. At large $U$, this transition occurs in the doublon band, while the singlon band is AL. Averaging was performed over 2000 different disorder realizations, $L=12$ sites, $N_\uparrow=6$ and $N_\downarrow=1$.}\label{fig:spectra}
\end{figure*}

{\it Analysis of resonances for a mobile impurity -- }We first examine the conditions under which a mobile impurity may induce resonances that involve nearby localized particles. We argue that for sufficiently weak disorder the impurity can propagate, thermalizing the entire system. We consider the Fermi-Hubbard model:
\begin{equation}\label{eq:fh}
\mathcal H_{\rm FH}=
\sum_{j,\sigma}\varepsilon_jn_{j,\sigma}+U\sum_jn_{j,\uparrow}n_{j,\downarrow}+t\sum_{j,\sigma}\Big[c^\dagger_{j,\sigma}c_{j+1,\sigma}+ \mbox{h.c.}\Big]\,,
\end{equation}
where the operators $c_{j,\sigma}$ annihilate spin-$\sigma$ fermions  on site $j$, $n_{j,\sigma}=c^\dagger_{j,\sigma}c_{j,\sigma}$, and the on-site energies $\varepsilon_j$ are uniformly distributed in the energy box $[-W,W$]. As an impurity, we introduce a single spin-down fermion, $N_\downarrow=1$, into an environment of $N_\uparrow$ spin-up fermions (particles) on $L$ sites, with $\rho_ \uparrow=N_\uparrow/L$ being their density. 

In the non-interacting limit ($U=0$), on-site disorder induces  AL.  It is convenient to switch to the basis of single-particle localized orbitals described by operators $a_{l,\sigma}$, related to lattice operators via $c_{j,\sigma}=\sum_l\psi_l(j)a_{l,\sigma}$. Here eigenfunctions $\psi_l(j)\sim e^{-|j-l|/\xi}/\sqrt{\xi}$ are exponentially localized around site $l$. The localization length $\xi$ scales as $t^2/W^2$ 
 at $W\ll t$~\cite{thouless_relation_1972,czycholl_conductivity_1981,kappus_anomaly_1981}. The  Hamiltonian~\eqref{eq:fh} becomes 
\begin{equation}
\begin{split}
&\mathcal H_{\rm FH}=\sum_l \mathcal E_la^\dagger_{l,\sigma}a_{l,\sigma}+\\&
~~U\sum_{ {j,l,m}\atop {p,q}}\psi_l^*(j)\psi_m(j)\psi_p^*(j)\psi_q(j)a^\dagger_{l,\uparrow}a_{m,\uparrow}a^\dagger_{p,\downarrow}a_{q,\downarrow}\,,
\end{split}
\end{equation} 
where $\{\mathcal E_l\}$ are the eigenenergies of the states $a_{l,\sigma}$.

The interaction of the impurity with the particles may induce resonances between initially fully localized configurations. To estimate the resonance probability, we study the matrix elements of the interaction term between an initial state $|\psi\rangle$ in which the impurity occupies orbital $l$, while spin-up fermions randomly occupy $N_\uparrow$ orbitals, and a final state $|\psi'\rangle$ in which the impurity moved to an orbital $l'$, and one spin-up fermion moved from orbital $p$ to $p'$. The corresponding matrix element reads
\begin{equation}\label{eq:matel}
\langle \psi'|\mathcal H_{\rm FH}|\psi\rangle= U\sum_j\psi^*_{p'}(j)\psi_p(j)\psi^*_{l'}(j)\psi_l(j)\,.
\end{equation}
We focus on processes where $l,l',p,p'$ all lie within one localization length $\xi$; for such processes the matrix element is largest. Assuming that $\psi_l(j)$ are oscillating functions of amplitude $1/\sqrt\xi$ within the localization volume, the matrix element~\eqref{eq:matel} can be estimated as $V_{\rm typ}\sim U\xi^{-3/2}$. Further, we note that a given state $|\psi\rangle$ is connected to $n(\xi,\rho_\uparrow)\sim \rho_\uparrow (1-\rho_\uparrow) \xi^3$ states $|\psi'\rangle$, since any of the $\rho_\uparrow \xi$ spin-up fermions can be moved to any of the $(1-\rho_\uparrow)\xi$ empty orbitals, while $l'$ can be chosen in $\sim \xi$ ways. The corresponding energy mismatch is a random quantity with a variance of order $\delta E\simeq 2W$. Thus, the level spacing for the rearrangement process can be estimated as $\delta\varepsilon=\delta E/n(\xi,\rho_\uparrow)$. The resonance condition, $V_{\rm typ}\gtrsim \delta\varepsilon$, then yields: 
\begin{equation}\label{eq:transitionU}
W<W_{\rm C}, \;\; 2W_{\rm C}= U\rho_\uparrow(1-\rho_\uparrow)\xi^{3/2}(W)\,.
\end{equation}
Thus, a single impurity efficiently induces many-body resonances in an initially localized system for $W<W_{\rm C}$. $W_{\rm C}$ is strongly sensitive to the precise dependence of the localization length $\xi$ on the disorder strength $W$, and on the spin-up density $\rho_\uparrow$. In particular, $W_{\rm C}$ is the largest for $\rho_\uparrow\approx1/2$, and it is zero in the $\rho_\uparrow\rightarrow0$ and $\rho_\uparrow\rightarrow1$ limits, signalling localization. In both limits,  Eq.~\eqref{eq:fh} maps onto a tight-binding model for a single spin-down  with random on-site potential, subject to AL.

\begin{figure}[t]
\includegraphics[width=.95\columnwidth]{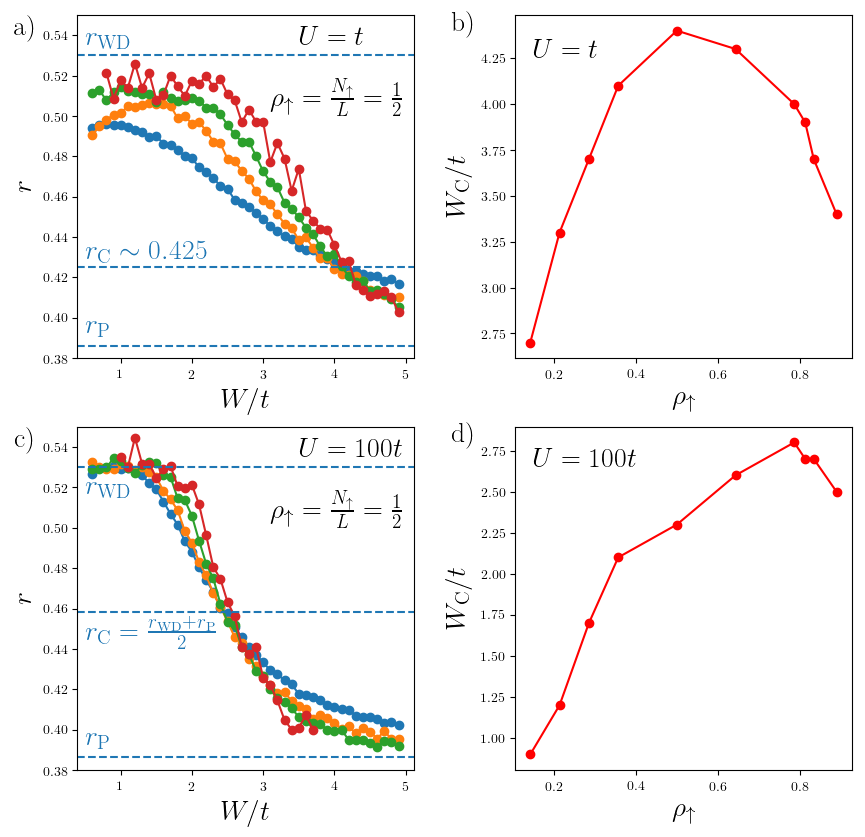}
\caption{a) Finite-size scaling of the averaged parameter $r$  as function of disorder strength $W/t$, for $U=t$.  {b)} Critical disorder $W_{\rm C}$, at which $r=r_{\rm C}$,  as a function of $\rho_\uparrow=N_\uparrow/L
$. The curve is consistent with the estimate~\eqref{eq:transitionU}, valid  in the thermodynamic limit. Disorder averages were performed over 4000, 2000, 1000 realizations for $L=8,\,10$ and 12. For $L=14,\,16$ and $18$ averaging is performed over 100 to 1000  realizations,  depending on $N_\uparrow$. c,d) Same as above, but for $U=100t$ and energies in the middle of the many-body band involving one doublon.  }\label{fig:transition}
\end{figure}

{\it Numerical calculations --  }We support the above considerations with ED numerical results presented in Figs.~\ref{fig:spectra}, \ref{fig:transition}, for the Fermi-Hubbard model~\eqref{eq:fh}. We first focus on moderate interactions ($U=t$), and the large-$U$ limit is discussed below.  In this case a clear transition from Wigner-Dyson (WD) to Poissonian (P) spectral statistics is observed as disorder is increased, in which the ratio of consecutive level spacings $r_n=\mbox{min}(\delta_n,\delta_{n+1})/\mbox{max}(\delta_n,\delta_{n+1})$ with $\delta_n=E_n-E_{n-1}$  goes from $r_{\rm WD}\sim0.53$ to $r_{\rm P}\sim 0.39$~\cite{oganesyan_localization_2007,atas_distribution_2013}.  Fig.~\ref{fig:transition}a reports finite-size scaling of the WD-P spectral transition, for $U=t$, suggesting persistence of thermalization induced by a single impurity in the thermodynamic limit.

Furthermore, we studied the effect of changing the spin-up density $\rho_\uparrow$ on the WD-P spectral transition (see  Fig.~\ref{fig:transition}c, and also SM~\footnotemark[1]). We estimate the critical disorder strength $W_{\rm C}$ as the crossing point of the $r$-parameter curves at different system sizes for $\rho_\uparrow=1/2$, which occurs at $r_{\rm C}\simeq0.425$, see Fig.~\ref{fig:transition}a. We assume the same $r_{\rm C}$ for arbitrary $\rho_\uparrow$. This yields the behavior of $W_{\rm C}(\rho_\uparrow)$ plotted in Fig.~\ref{fig:transition}b, which is in a qualitative agreement with the estimate~\eqref{eq:transitionU}.

\begin{figure}[t]
\includegraphics[width=.95\columnwidth]{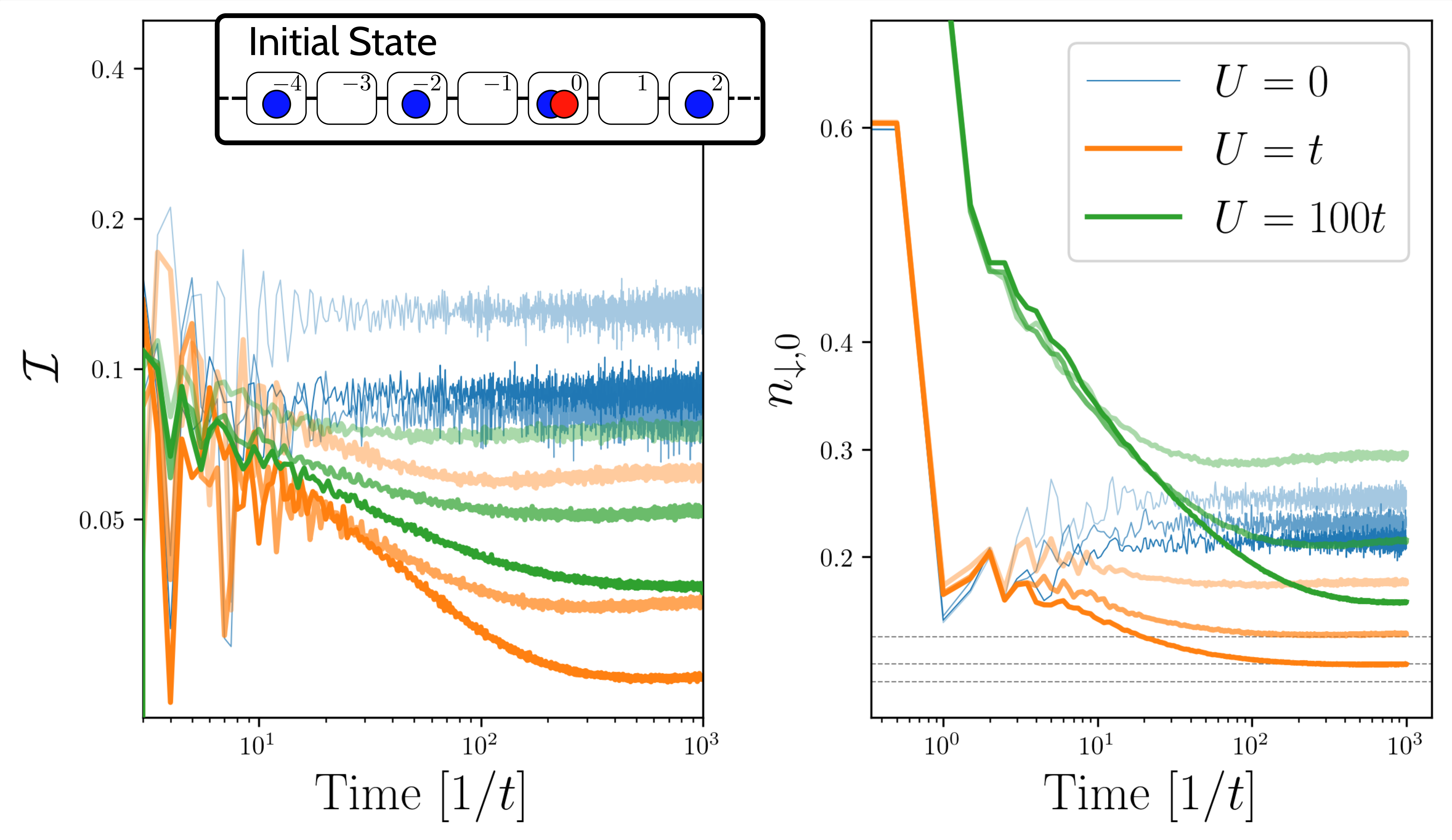}
\caption{ Left: Time evolution of the imbalance $\mathcal I$  for the initial state sketched in the inset. Different interaction strengths are considered ($U=0,t,100t$), and periodic boundary conditions are chosen. Right: The evolution of the local spin-down density at site zero, $n_{\downarrow,0}$.  Simulations are performed for $L=8,10,12$ (light to dark lines), $N_\uparrow =L/2$, $W=1.5t$ and averaged over 1000 disorder realizations. The horizontal gray lines correspond to $1/L$. }\label{fig:diffusion}
\end{figure}

{\it Dynamical properties --} Next, we focus on dynamical properties and study impurity propagation and its effect on the initially localized spin-up particles. We consider a quantum quench protocol, with an initial state chosen to be a product state of a spin-down particle located on a lattice site $0$, and a density wave, with a spin-up fermion occupying every second lattice site, Fig.~\ref{fig:diffusion}. We study the evolution of the spin-up imbalance $\mathcal I=(N_{\rm \uparrow,e}-N_{\rm \uparrow,o})/(N_{\rm \uparrow,e}+N_{\rm \uparrow,o})$, $N_{\rm \uparrow,e/o}$ being the occupation of even/odd sites (see experiments~\cite{schreiber_observation_2015,boll_spin-_2016}), as well as the averaged evolution of the local spin-up (impurity) occupation  $\langle n_{\downarrow,0}\rangle$. 

The results for $W=1.5t$, which lies below $W_{\rm C}$ for $U=t$, and $\rho_\uparrow=1/2$, are illustrated in Fig.~\ref{fig:diffusion}. In the non-interacting case ($U=0t$), neither of the quantities $\langle n_{\downarrow,0}\rangle, \mathcal{I}$ thermalizes because of AL, saturating to a value that depends weakly on the system size. At $U\neq0$, both quantities show a more pronounced decay as the system size is increased.  In particular, the relaxation of $\langle n_{\downarrow,0}\rangle$ towards $1/L$ signals  uniform spreading of the mobile impurity across the entire system. The suppression of the imbalance $\mathcal I$ is correlated in time with the impurity spreading across the system. Measurements of the imbalance decay and impurity position can be readily performed in ongoing experiments~\cite{schreiber_observation_2015,boll_spin-_2016,chiaro_growth_2019} and would provide a smoking-gun signature of the nucleation mechanism.


{\it Mesoscopic nucleation and doublons in the $U\rightarrow\infty$ limit -- }  Figure~\ref{fig:spectra}c shows that the condition $U\gg N_\uparrow \cdot\mbox{max}[t,W]$ induces a mesoscopic gap between two different many-body bands with distinct spectral properties. Such two bands are distinguished by the number of doublons present in the system.  A doublon is formed when two fermions of opposite spin occupy  the same site, with an energy cost $U$.  In the limit $U\to\infty$, doublons are stable excitations at $t\neq 0$, as the energy $U$ released by doublon decay cannot be absorbed by bands without doublons~\cite{strohmaier_observation_2010}.  In this limit, the low-energy ``singlon'' band, in which the unique spin-down fermion cannot occupy sites hosting spin-up fermions, is localized, as indicated by the Poisson level statistics. Similar to bosons in the Tonks-Girardeau gas~\cite{tonks_complete_1936,girardeau_relationship_1960,lieb_exact_1963,michal_finite-temperature_2016,aleiner_finite-temperature_2010} (see SM for details~\footnote{See Supplementary Material.}), the $U\rightarrow\infty$ limit induces a Pauli-like exclusion between spin-up and -down fermions, which then behave as indistinguishable, Anderson-localized free fermions, see Fig.~\ref{fig:spectra}c.

\begin{figure}[t]
\includegraphics[width=.95\columnwidth]{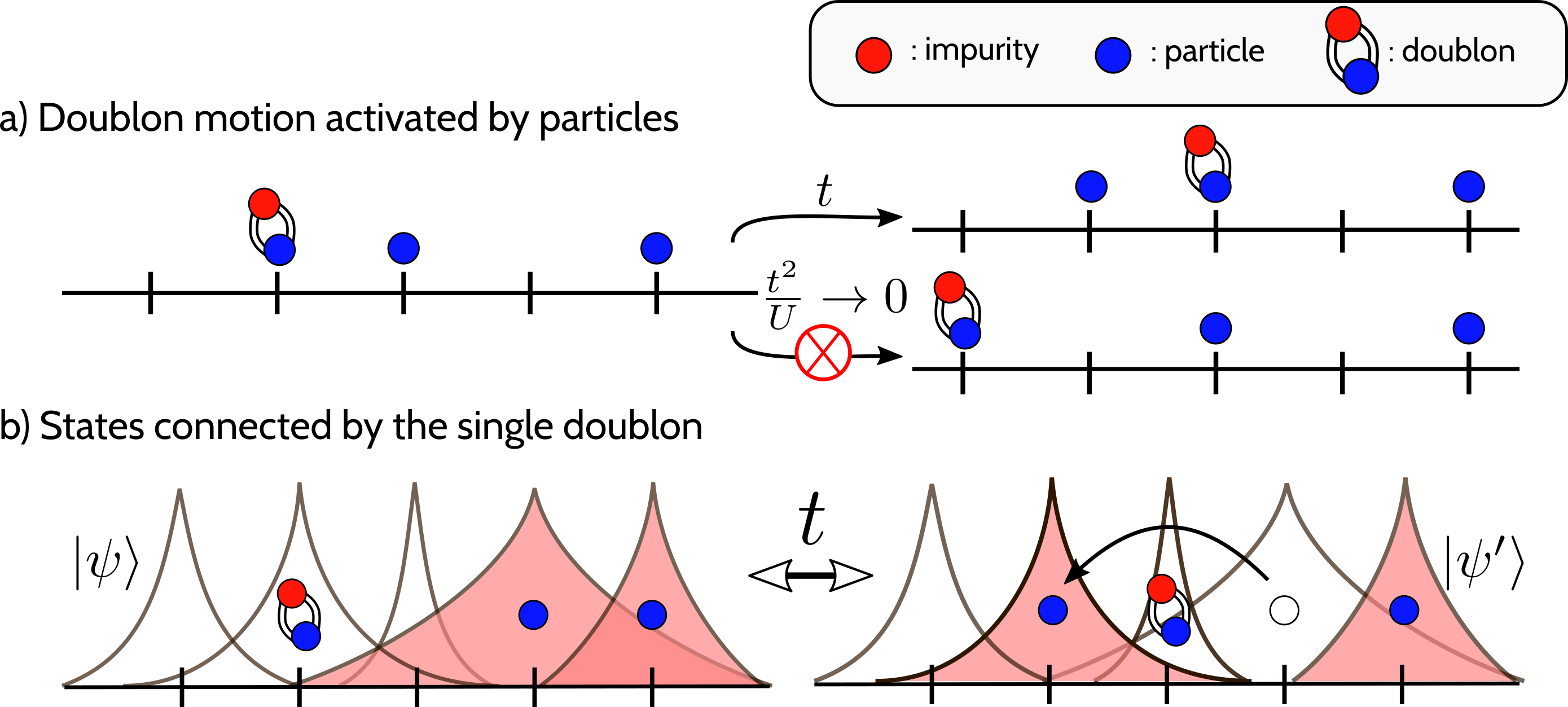}
\caption{a) Effective correlated hopping processes of doublon and particles. b) Typical states connected to leading order in $t$. Doublon-hopping is accompanied in the change of occupation of localized orbitals and the displacement of impenetrable holes. }\label{fig:matel_doublon}
\end{figure}

A pronounced WD-P transition as a function of disorder strength is observed in the single doublon sector for $U=100t$. Figure~\ref{fig:transition}c shows that the step in $r$ is sharper than the one observed for moderate interactions ($U=t$); moreover, $r$ reaches $r_{\rm WD}$ for accessible sizes $L$ on the ergodic side of the step. Figure~\ref{fig:transition}d shows that the WD-P transition depends on the spin-up density $\rho_\uparrow$, but $W_{\rm C}(\rho_\uparrow)$ is less symmetric around $\rho_\uparrow=1/2$. In the SM~\footnotemark[1], a similar transition is  demonstrated for indistinguishable bosons described by the Bose-Hubbard model when a single doublon is stabilized by large $U\rightarrow\infty$. 

The emergence of  WD statistics shows  that, at sufficiently weak disorder, a single doublon can efficiently induce many-body resonances. The underlying processes are illustrated in Fig.~\ref{fig:matel_doublon} and are described in detail in  the SM~\footnotemark[1]. In the $U\rightarrow\infty$ limit, the doublon hopping is impossible in the absence of particles nearby. In this case,  it involves a virtual transition through states without doublons, and therefore has an amplitude $\mathcal O(t^2/U)$ that vanishes for $U\rightarrow \infty$.  On the other hand, the hopping of the unique spin-down fermion on a nearby site already  occupied by a spin-up particle is allowed. It has an amplitude $t$ and conserves the doublon number. This constrained dynamics, summarized in Fig.~\ref{fig:matel_doublon}a, may induce resonances between localized configurations, as those shown in Fig.~\ref{fig:matel_doublon}b, favoring ergodicity for weak enough disorder. The analysis of the resonances induced by such correlated hopping is carried out in the SM~\footnotemark[1] to the leading order in $t$. It is similar to the argument underlying Eq.~\eqref{eq:transitionU}, with the difference that $\delta E\sim3W$ (the impurity and two particle are displaced), and the number of connected states is estimated as $\sim \rho_\uparrow (1-\rho_\uparrow)\xi^2$ (the doublon moves by one site and only one particle-hole pair is rearranged in a localization volume $\xi$). This yields~\footnotemark[1] 
\begin{equation}\label{eq:transition}
3W\leq t\rho_\uparrow(1-\rho_\uparrow)\sqrt{\xi(W)}\,.
\end{equation}
Thus, the hopping amplitude  $t$ substitutes the interaction strength $U$ in Eq.~\eqref{eq:transitionU}, and  the  existence of resonances is controlled by a single parameter $W/t$, the only one left in the $U\rightarrow\infty$ limit.  The condition~\eqref{eq:transition} explains the observation of the WD-P transition in Figs.~\ref{fig:spectra}c and~\ref{fig:transition}c, which is  a consequence of the fact that the doublon is  able to induce many-body resonances if a  finite density of particles is present nearby. We attribute the observed deviations from particle-hole symmetric scaling $\propto \rho_\uparrow(1-\rho_\uparrow)$  to finite-size effects, observing that the numerical results drift towards the dependence  predicted by Eq.~\eqref{eq:transition} as the system size is increased, see SM~\footnotemark[1]. 

Nevertheless, a single doublon cannot spread over the entire system in the
thermodynamic limit. In a strict $U\rightarrow\infty$ and $d=1$ limit, doublons cannot cross the holes present in the system. As shown schematically in Fig.~\ref{fig:matel_doublon}b, doublons and particles exchange positions through processes of amplitude $t$, but the same process triggers the accumulation of holes on one side of the doublon and will eventually block its propagation. 
This effect becomes dominant after repeated hopping events of order $t$, exchanging particle and doublons, and is apparent in the real-time evolutions shown in Fig.~\ref{fig:diffusion}. In particular, the stationary long-time value of $\langle n_{\downarrow,0}\rangle$ at $U=100t$ shows no convergence towards $1/L$.  
In the SM~\footnotemark[1],  we compare the doublon spreading for periodic and open boundary conditions, showing that it is significantly suppressed in the latter case. The reason is that, for open boundary conditions, the number of empty sites on the left and on the right of the initial doublon position is a {\it conserved} quantity in the $U\rightarrow\infty$ limit, which quenches the doublon propagation, even in the absence of many-body localization.  The doublon is thus able to take advantage of particles nearby to propagate on mesoscopic length scales which are much larger than the single particle localization length and the systems sizes attained by our numerical simulations. 


It is important to stress that the above considerations do not generalize to the case with moderate interactions $U\sim t$, in which case there is no mechanism blocking the propagation of the interacting impurity at weak disorder across the entire system.


{\it Conclusions -- } In this work, we showed that a single impurity may thermalize an initially localized system at sufficiently weak disorder; such system is therefore a ``supercooled insulator" that stays localized only until a suitable ergodic seed is introduced. Crucially, the associated thermalization time is expected to  scale as a power-law in system size, while static bubbles of Ref.~\cite{de_roeck_stability_2017} lead to much slower, exponential in size relaxation.

This effect  can be observed with current experimental capabilities, e.g. by studying dynamical signatures and impurity-induced decay imbalance. In the future work, it would be interesting to extend the analysis to the case of a finite impurity density. This may provide an insight into the results of a recent experiment~\cite{rubio-abadal_many-body_2019}, which reported MBL-thermal transition as a function of impurity density. Finally, several open questions regarding the doublon dynamics in the $U\to\infty $ limit remain; in particular, doublon dynamics in $d>1$ is expected to be qualitatively different from $d=1$ case, since no additional conservation law exists. In particular, we expect that, for  $d>1$, the doublon surrounded by its ergodic cloud  will preserve its ability to propagate across the whole, initially localized, system.

{\it Acknowledgments ---}
%
M.F. is particularly indebted to Denis Basko and Anna Minguzzi for the past discussions at the origin of this work and acknowledges  Vadim Cheianov, Gabriel Lemari\'e, Thierry Giamarchi, Markus M\"uller, Wojciech De Roeck, Felix von Oppen and Anne-Maria Visuri for useful comments. P. W. B. acknowledges support
by project A03 of the CRC TR 183 of the Deutsche Forschungsgemeinschaft
DFG.  D.A. thanks SNSF for support under Division II. M.F. also acknowledges support from the FNS/SNF Ambizione Grant PZ00P2\_174038.

\bibliographystyle{apsrev4-1}
\bibliography{biblio,bib_2}

\end{document}



\title{Supplementary Material of ``Nucleation of ergodicity by a single mobile impurity in supercooled insulators''}

\author{Ulrich Krause$^{1}$}
\author{Th\'eo Pellegrin$^{2}$}
\author{Piet W. Brouwer$^{1}$}
\author{Dmitry A. Abanin$^{3}$}
\author{Michele Filippone$^{2}$}

\affiliation{$^1$Dahlem Center for Complex Quantum Systems and Institut f\"ur Theoretische Physik, Freie Universit\"at Berlin, Arnimallee 14, 14195 Berlin, Germany}

\affiliation{$^2$Department of Quantum Matter Physics and $^3$Department of Theoretical Physics, University of Geneva, 24 Quai Ernest-Ansermet, CH-1211 Geneva, Switzerland}


\begin{abstract}
In this Supplemental Material, we provide additional and detailed information concerning the numerical calculations presented in the main text for the Fermi-Hubbard model. We then compare the spreading of the mobile impurity by assuming periodic and open boundary conditions.  We also  show how the ergodic bubble generation by single doublons equally emerges in bosonic systems, by carrying an analog numerical study on the Bose-Hubbard model. We conclude by carrying out the analysis of resonances in the presence of a doublon, leading to Eq.~(5) in the main text.   
\end{abstract}


\maketitle
\setcounter{equation}{0}
\setcounter{figure}{0}
\setcounter{table}{0}
\setcounter{page}{1}
\makeatletter
\renewcommand{\theequation}{S\arabic{equation}}
\renewcommand{\thefigure}{S\arabic{figure}}
\renewcommand{\bibnumfmt}[1]{[S#1]}
\renewcommand{\citenumfont}[1]{S#1}

\onecolumngrid

\newpage

\section{Additional details on numerical density- and finite-size scaling}
In this section, we provide additional plots showing how Wigner-Dyson statistics is induced by the presence of a single impurity for different spin-up densities $\rho_\uparrow=N_\uparrow/L$ and system sizes $L$.  Figures~\ref{fig:finiteU1} and~\ref{fig:finite} report analogous plots to Fig.~1 in the main text. We focus   on $U=t$ and $U=100t$, by considering different system sizes and different particle densities $\rho_\uparrow$. Notice that we inverted the axes and switched to a different scale for the vertical axis. Instead of the energy of the state, we consider its position in the many-body spectrum. This choice has limited physical meaning, but it allows for instance to understand how the clear lines appearing in Fig.~1a for weak disorder in the ergodic region are an artifact of ``folding'' the spectrum. Additionally one can also notice that such lines are a finite-size effects, much less apparent for  $L=14$.

\newcommand{\cst}[0]{.22}

\begin{figure}[h!]
\includegraphics[height=\cst\textwidth]{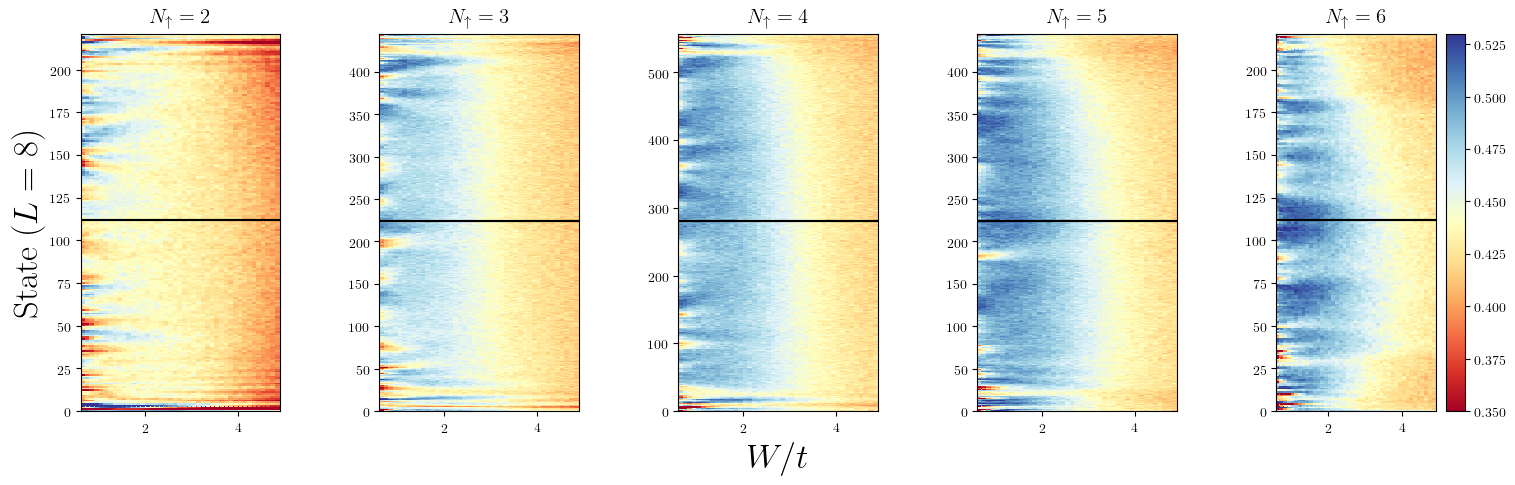}
\includegraphics[height=\cst\textwidth]{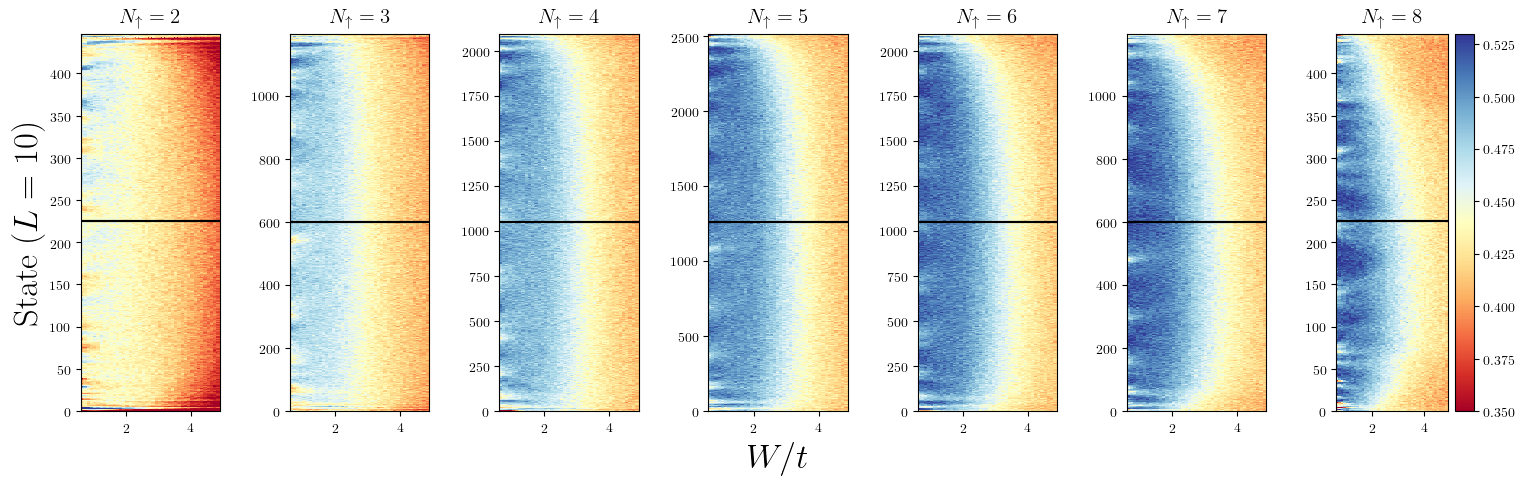}
\includegraphics[height=\cst\textwidth]{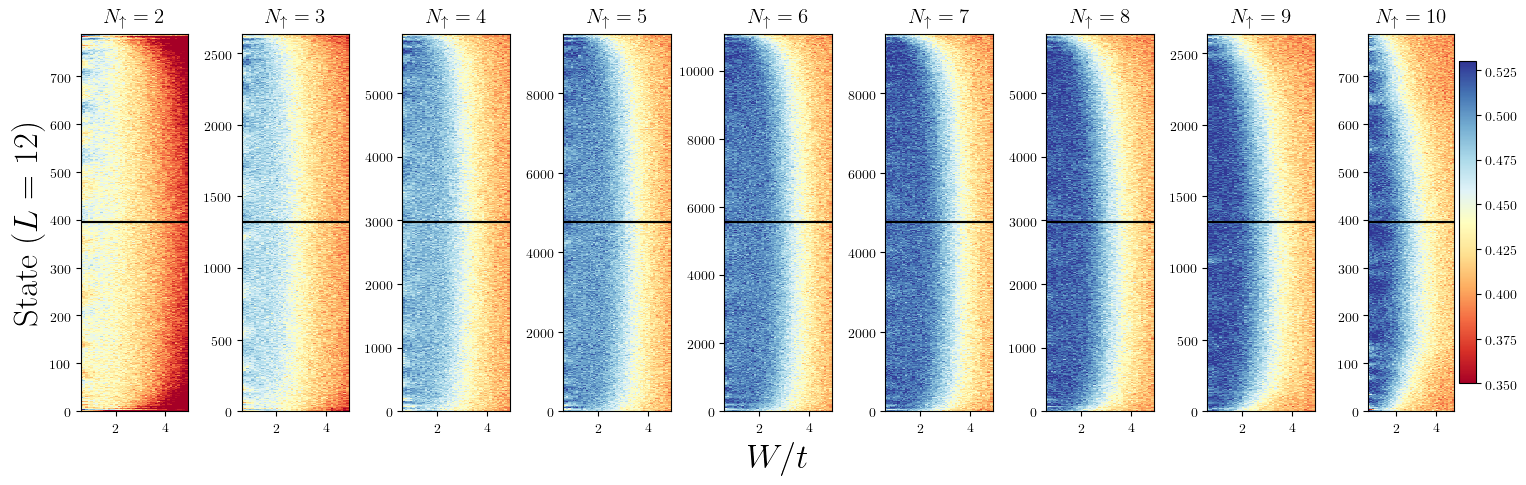}
\includegraphics[height=\cst\textwidth]{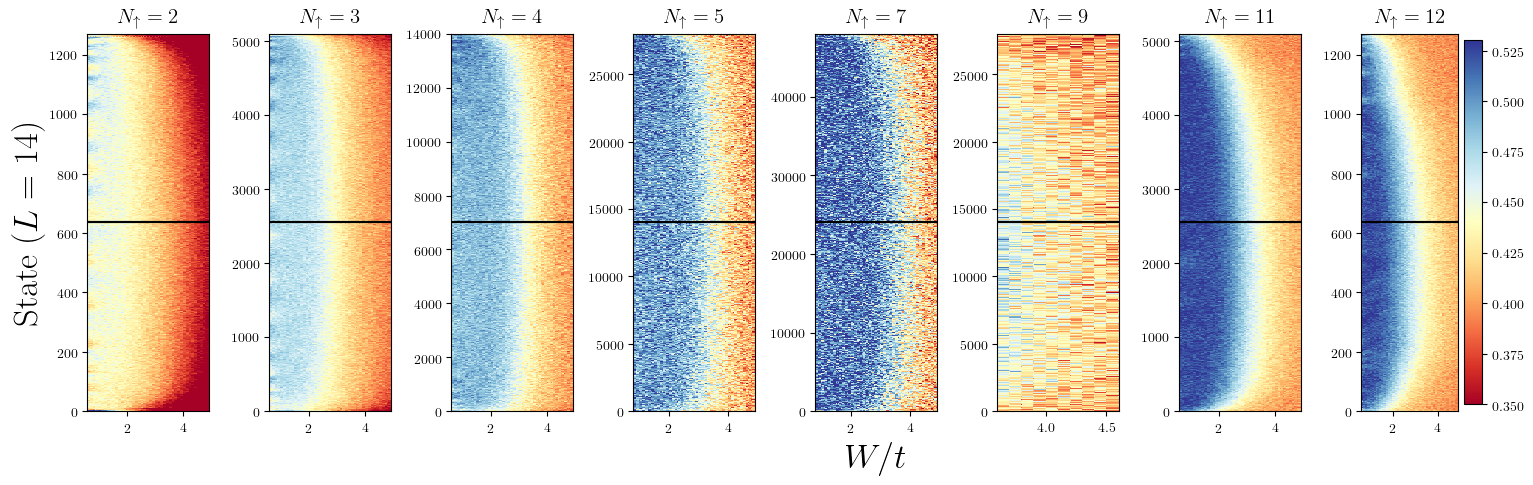}
\caption{Density plot of the averaged gap-ratio $r$ as function of disorder strength $W/t$ and state label (see text), for moderate interaction strength $U=t$. The black horizontal lines signal the state in the middle many-body bands and give the cuts along which the lines in Fig.~\ref{fig:rhosU1} are taken.  Disorder averages were performed over 4000, 2000, 1000 realizations for $L=8,\,10$ and 12. For $L=14$ averages over disorder range from 1000 to 100 different disorder realizations depending on $N_\uparrow$.}\label{fig:finiteU1}
\end{figure}

Additionally, in Fig.~\ref{fig:finite}, considering the position of the states instead of their energy does not make apparent  the ``mobility gap'' in energy between doublon and singlon bands for $U=100t$, which is instead clearly apparent in Fig. 1c. Nevertheless, it allows to clearly distinguish many-body bands with and without doublons and better appreciate the transition from Wigner-Dyson to Poisson level-spacing statistics. We remind that, for the Fermi-Hubbard model with $N_\uparrow$ and $N_\downarrow$ spin-up and -down fermions on $L$ sites, the Hilbert space dimension is $\binom{L}{N_\uparrow}\times\binom{L}{N_\downarrow}$. For the specific case in which $N_\downarrow=1$, the singlon and doublon bands feature $L\times\binom{L-1}{N_\uparrow}$ and $L\times\binom{L-1}{N_\uparrow-1}$ states respectively.  

\begin{figure}
\includegraphics[height=\cst\textwidth]{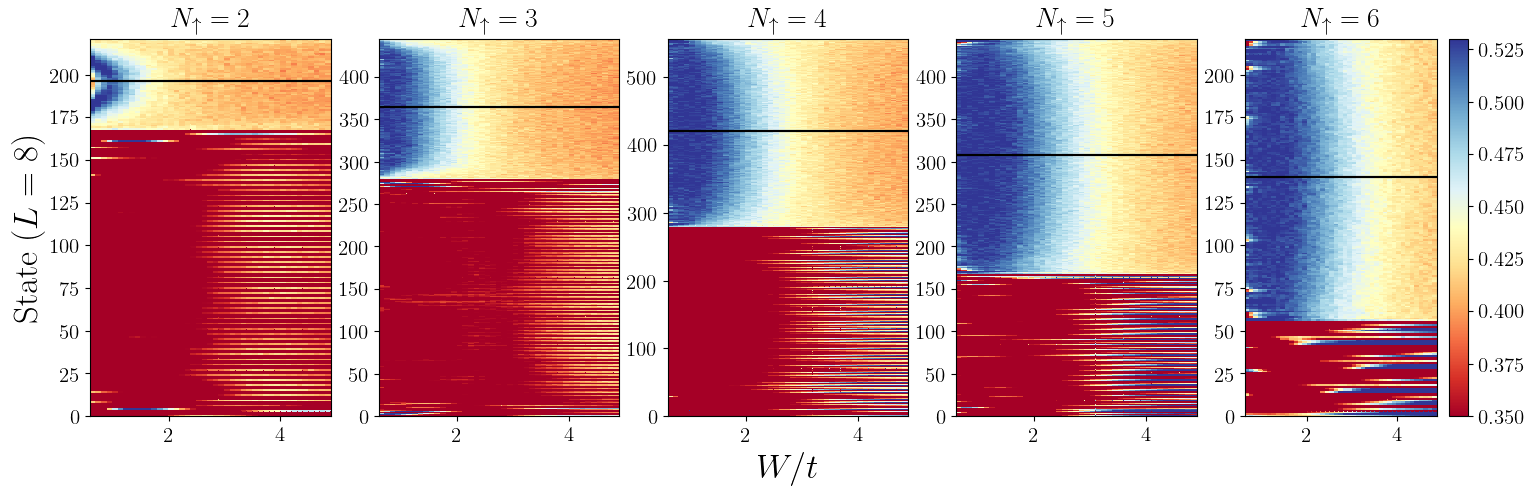}
\includegraphics[height=\cst\textwidth]{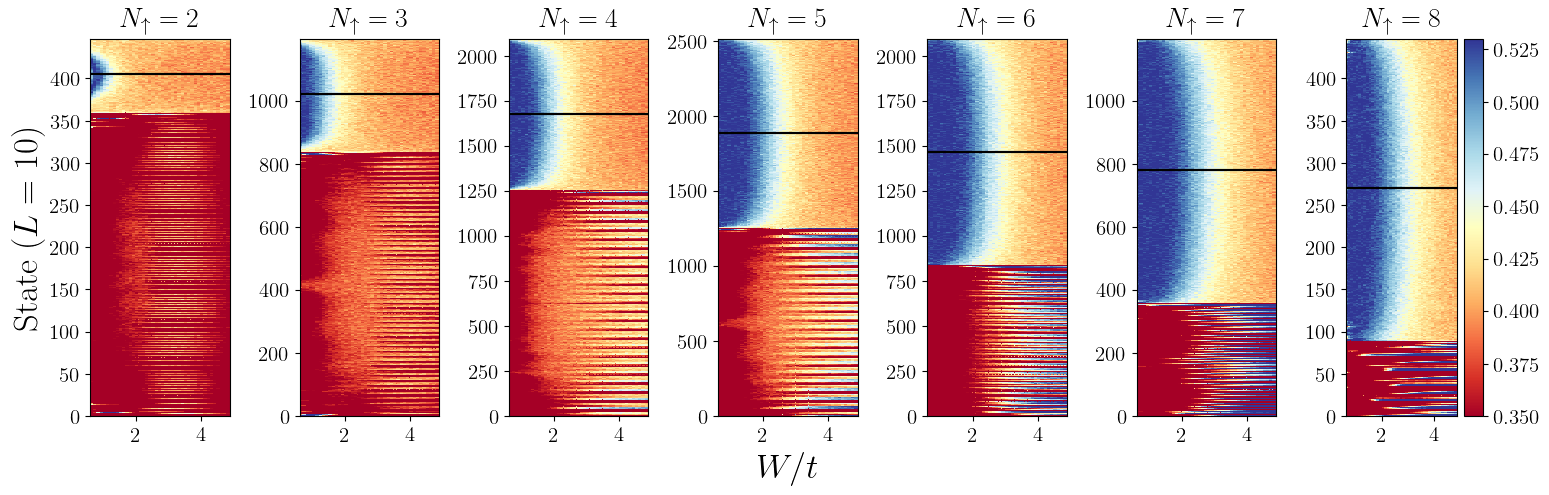}
\includegraphics[height=\cst\textwidth]{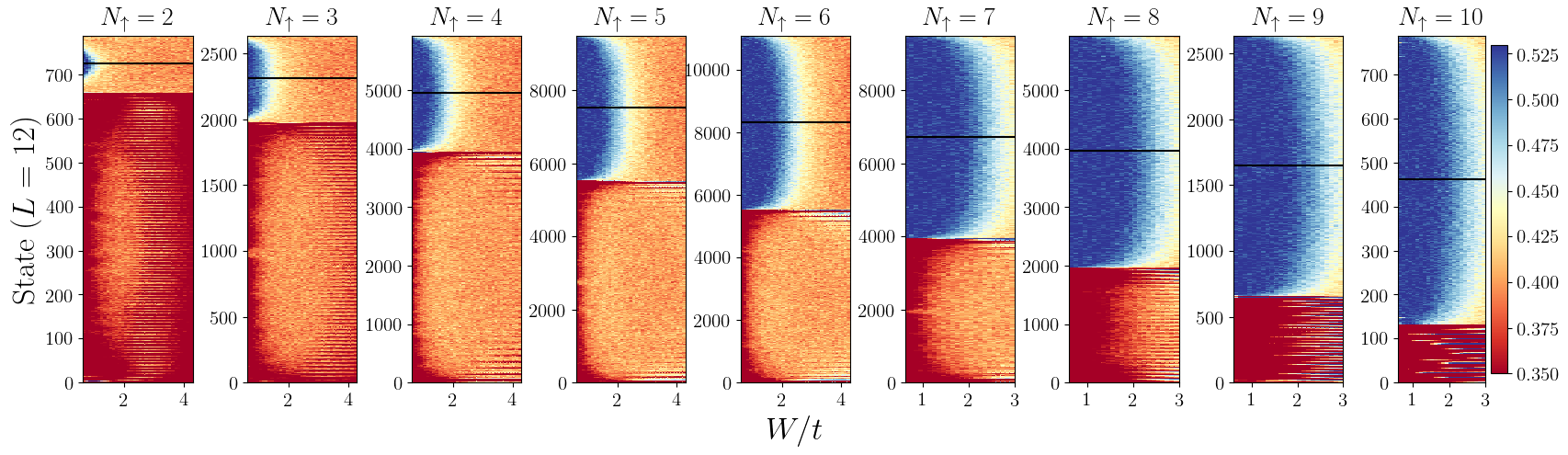}
\includegraphics[height=\cst\textwidth]{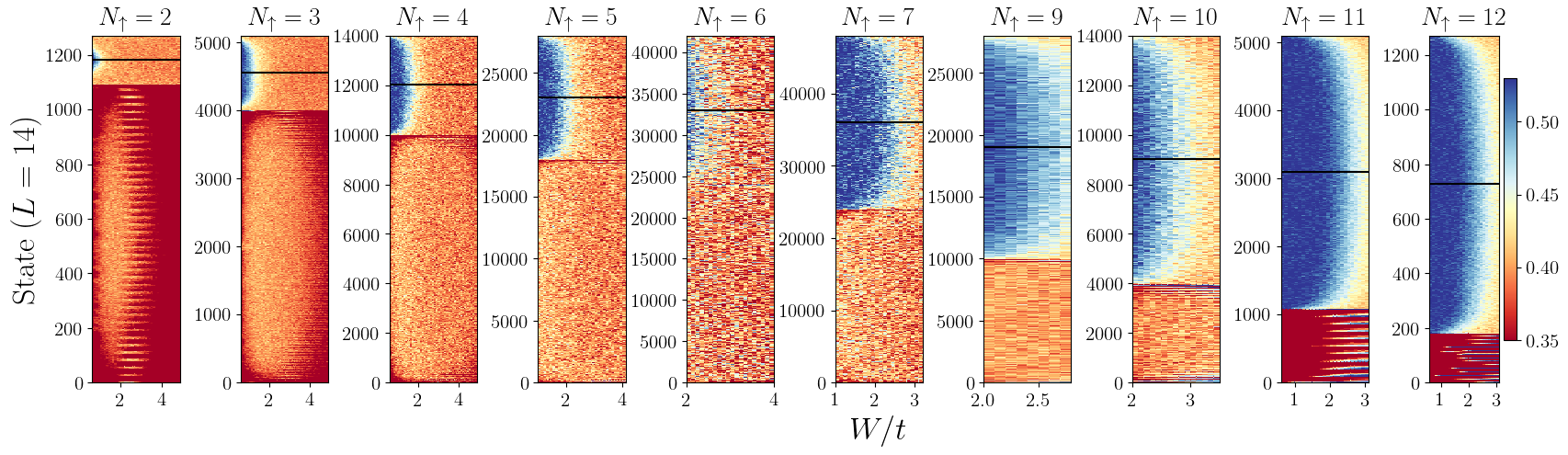}
\caption{Same as Fig.~\ref{fig:finiteU1} for strong interactions $U=100t$, clearly showing the appearance of ergodicity only in the bands hosting doublons. The black horizontal lines signal the state in the middle of the doublon bands and give the cuts along which the lines in Fig.~\ref{fig:rhos} are taken. The red regions appearing on the bottom correspond to the AL bands with only singlons. All features there are artifacts of the numerical averaging procedure.}\label{fig:finite}
\end{figure}

Figures~\ref{fig:rhosU1} and~\ref{fig:rhos} provide details concerning the derivation of the critical disorder strength $W_{\rm C}$ as a function of the spin-up density $\rho_\uparrow$ shown in Fig.~2.b-d in the main text, for different system sizes.

\newcommand{\cstn}[0]{.44}

\begin{figure}[h]
\includegraphics[width=\cstn\textwidth]{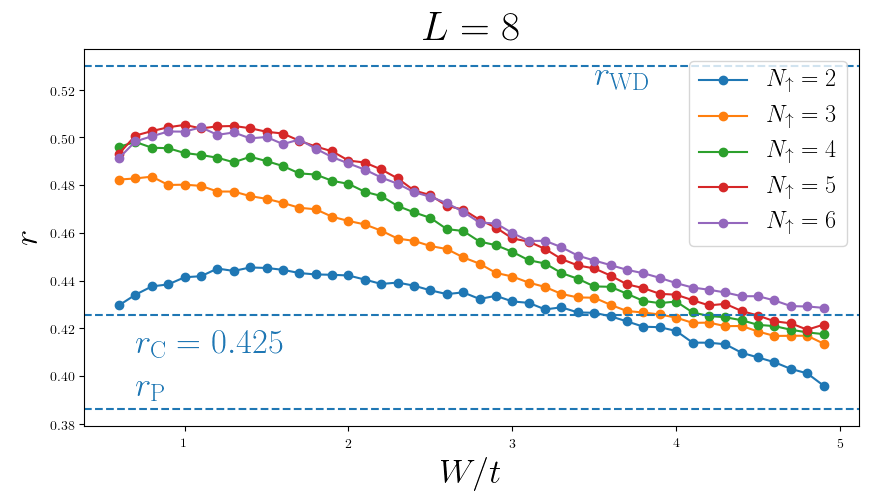}
\includegraphics[width=\cstn\textwidth]{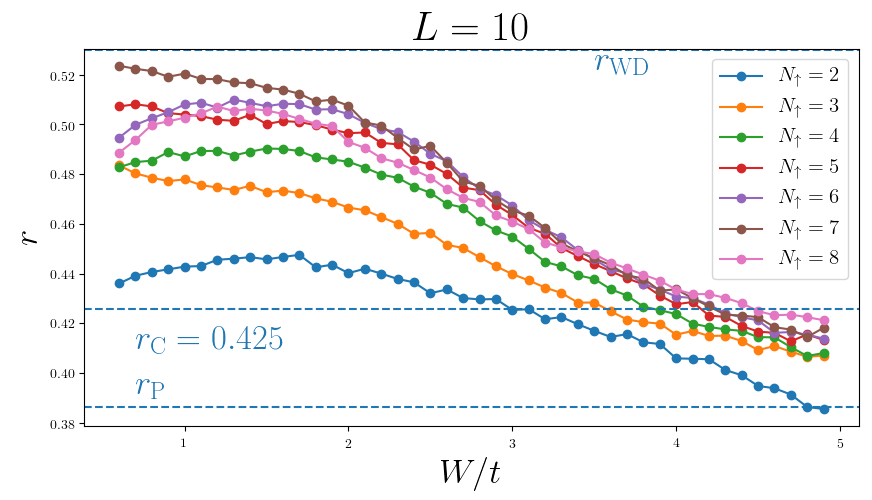}
\includegraphics[width=\cstn\textwidth]{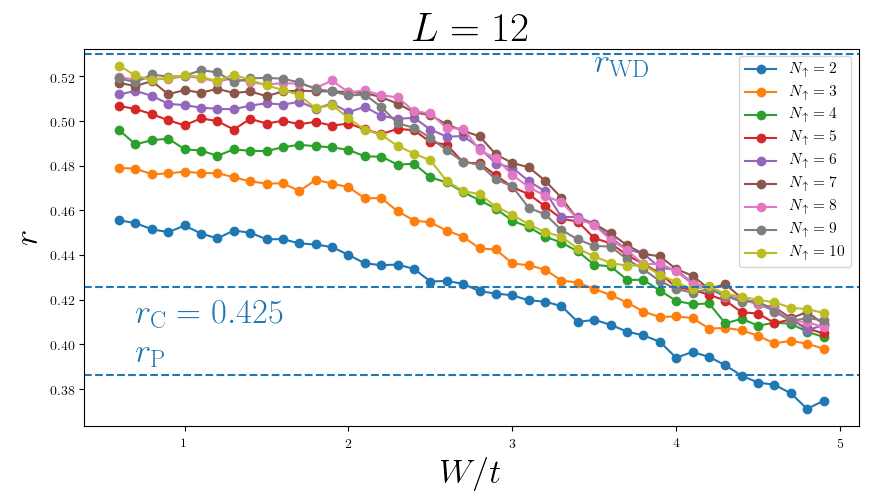}
\includegraphics[width=\cstn\textwidth]{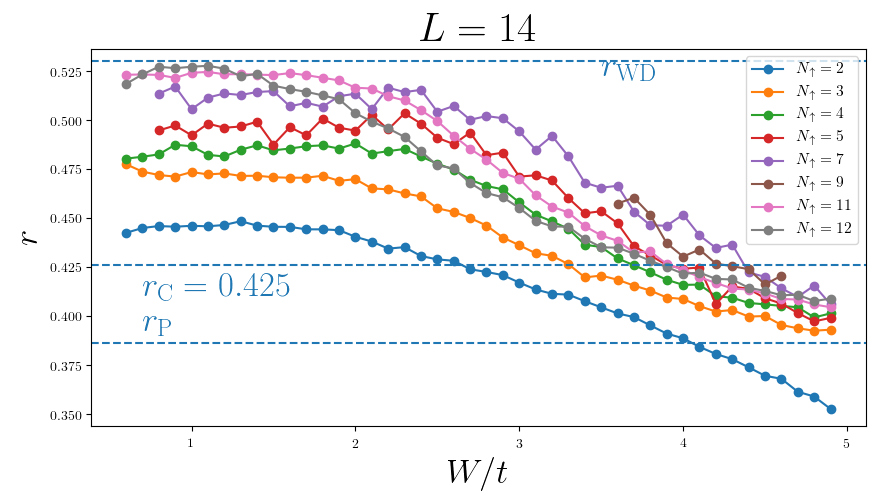}
\caption{Averaged gap-ratio $r$ along the black lines drawn in Fig.~\ref{fig:finiteU1}. Different plots correspond to different system size $L$.}\label{fig:rhosU1}
\end{figure}

\begin{figure}[h]
\includegraphics[width=\cstn\textwidth]{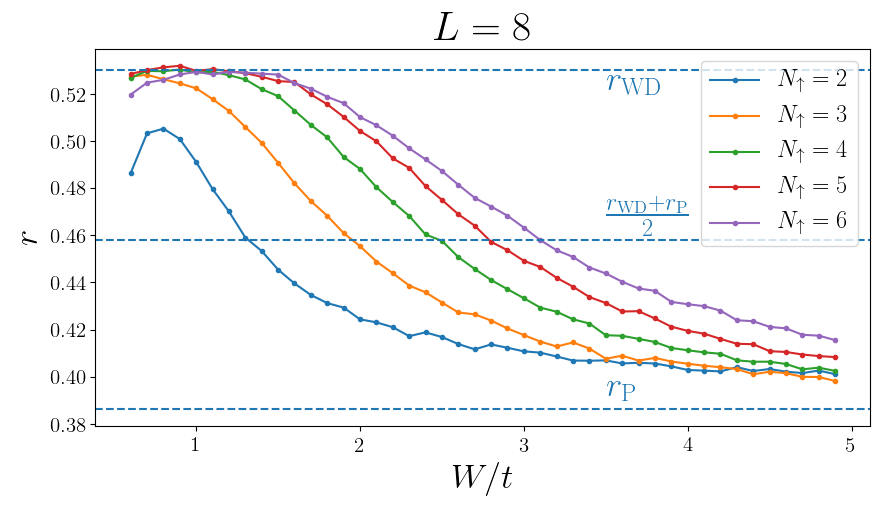}
\includegraphics[width=\cstn\textwidth]{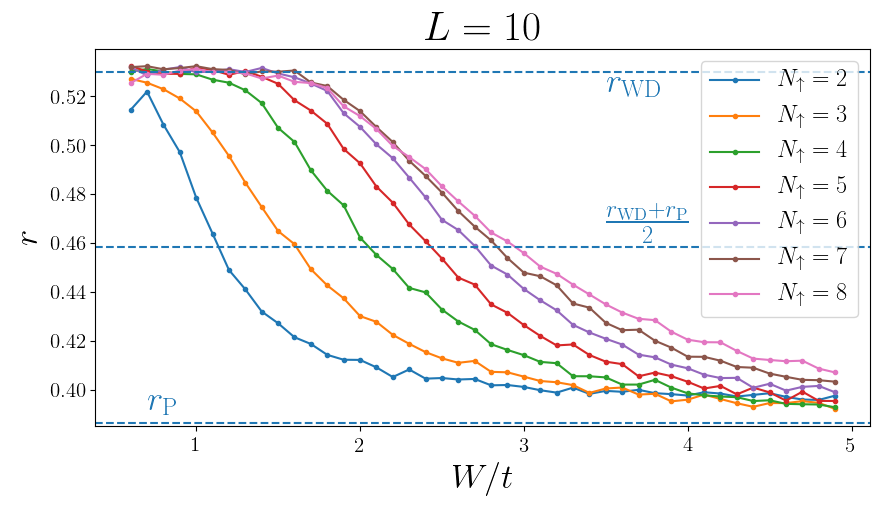}
\includegraphics[width=\cstn\textwidth]{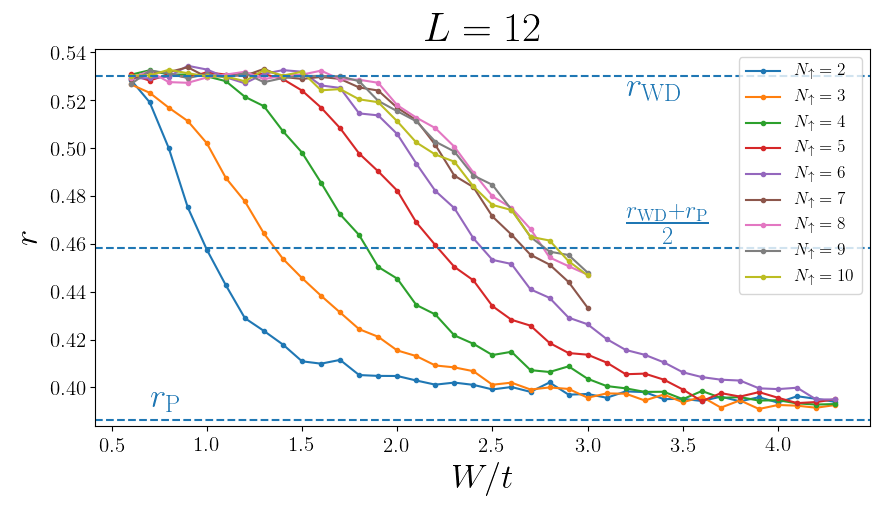}
\includegraphics[width=\cstn\textwidth]{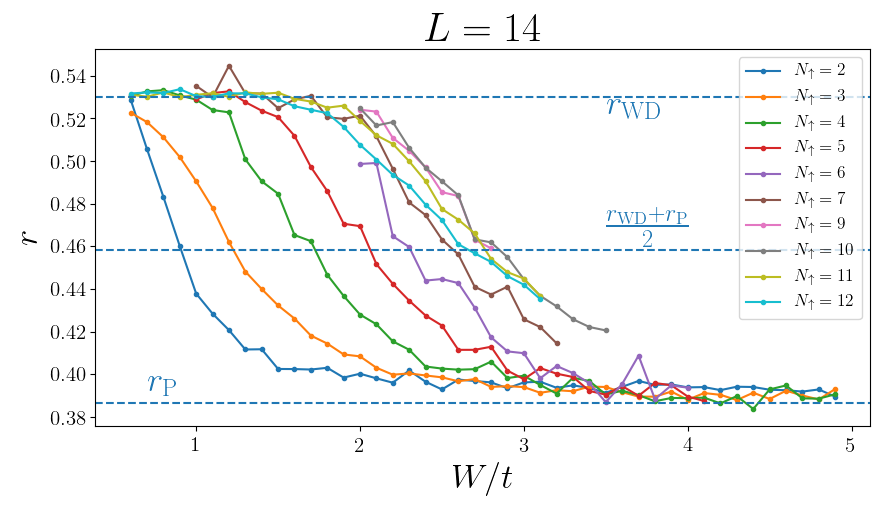}
\caption{Same as Fig.~\ref{fig:rhosU1} for $U=100t$.}\label{fig:rhos}
\end{figure}

\begin{figure}[h]
\includegraphics[width=.6\textwidth]{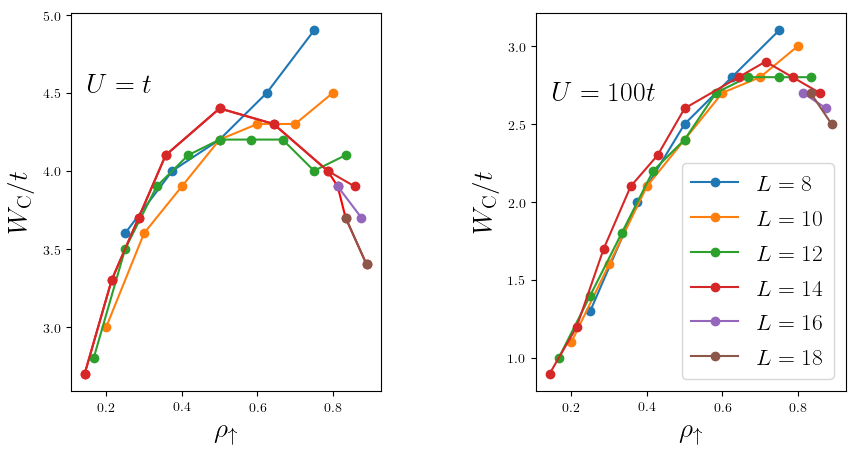}
\caption{Finite-size scaling of $W_{\rm C}$ derived from the numerical data presented in Figs.~\ref{fig:rhosU1} and~\ref{fig:rhos}, for $U=t$ and $U=100t$ respectively.  }\label{fig:wc_L}
\end{figure}

In Fig.~\ref{fig:wc_L}, we provide also the finite-size scaling of the critical disorder strength $W_{\rm C}$ evaluated from the numerical data presented in Figs.~\ref{fig:rhosU1} and~\ref{fig:rhos}. Figure 2b-d, in the main text, have been constructed from Fig.~\ref{fig:wc_L}, by keeping the data available for the largest system size $L$ for a given density $\rho_\uparrow$. Figure~\ref{fig:wc_L} clearly shows that the numerical results drift towards the dependence $W_{\rm C}\propto\rho_\uparrow(1-\rho_\uparrow)$ predicted by Eqs.~(3) and~(4) presented in the main text. 

~
\newpage

\section{Sensitivity of the doublon spreading on boundary conditions}
\begin{figure}[b]
\includegraphics[width=\textwidth]{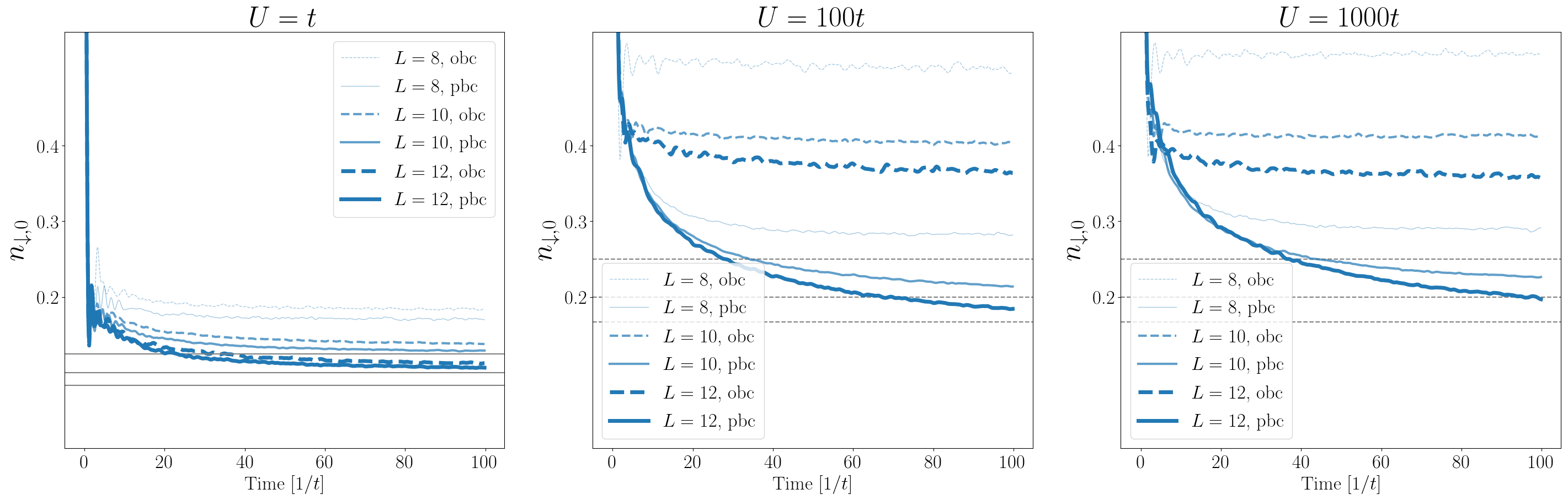}
\caption{Comparison of the evolution of the local spin-down occupation at site zero $n_{\downarrow,0}$ for periodic (solid lines) and open boundary conditions (dashed lines). The initial state is the same as the one considered for Fig.~3 in the main text. Increasing values of the interaction strength $U$ are considered from left to right ($t,100t,1000t$).  Simulations are performed for $L=8,10,12$ (light to dark lines), $N_\uparrow =L/2$, $W=1.5t$ and averaged over 200 disorder realizations. The horizontal lines in gray  correspond to $1/L$ (solid), and $2/L$ (dashed).  }\label{fig:evolveobc}
\end{figure}

As discussed in the main text, in the case of open boundary conditions (obc), the number of empty sites (holes) on the left and on the right of the doublon becomes a quasi-conserved quantity in the $U\gg t$ limit. As a consequence, the spreading of the ergodic cloud activated by the doublon is quenched by the presence of such ``incompressible'' holes. This phenomenon is illustrated in Fig.~\ref{fig:evolveobc}, in which we compare the evolution towards its stationary value of the spin-down density $n_{\downarrow,0}$ in the case of open and periodic boundary conditions (pbc). The spin-down fermion is initially placed at site $0$ with a singlon every two sites ($\rho_\uparrow=1/2$), as for the initial state in Fig. 3 in the main text. In the case of moderate interaction ($U=t$), the dynamics is almost insensitive to the boundary conditions and the stationary value of $n_{\downarrow,0}$ well approaches the ergodic value $1/L$. 

The situation is different when the interaction is increased and the doublon is stabilized. In that case, the evolution of $n_{\downarrow,0}$ is strongly sensitive to boundary conditions. In the case of pbc, the stationary value of $n_{\downarrow,0}$ approaches $2/L$, that is the ergodic value that would be obtained by subtracting to the system the sites occupied by holes. For obc instead, the stationary value of $n_{\downarrow,0}$ is quite different from this value, signaling the difficulty of the doublon in propagating across the system because of the presence of an additional  quasi-conserved quantity, namely the number of holes on the right and on the left of the doublon. 

\section{Ergodic bubble generation by single doublons in bosonic systems}
In this section we consider the Bose-Hubbard model, which is the bosonic version of the Fermi-Hubbard model we considered in  Eq.~(1) in the main text. It reads
\begin{equation}\label{eq:bh}
\mathcal H_{\rm BH}=t\sum_{j}\Big[b^\dagger_j b_{j+1}+b^\dagger_{j+1}b_j\Big]+U\sum_{j}n_j(n_j-1)+\sum_{j}\varepsilon_j n_j\,,
\end{equation}
in which the operators $b_{j,\sigma}$ annihilate bosons on site $j$, $n_{j,\sigma}=b^\dagger_{j}b_{j}$, and the onsite energies $\varepsilon_j$ are uniformly distributed in the energy window $[-W,W$]. Notice that for boson, doublon formation does not need to add any internal degree of freedom, like spin, as the bosonic statistics allows for the multiple occupation of a single site, thus permitting the formation of doublons, even  triplons and so forth. 

\begin{figure}[t]
\includegraphics[width=\textwidth]{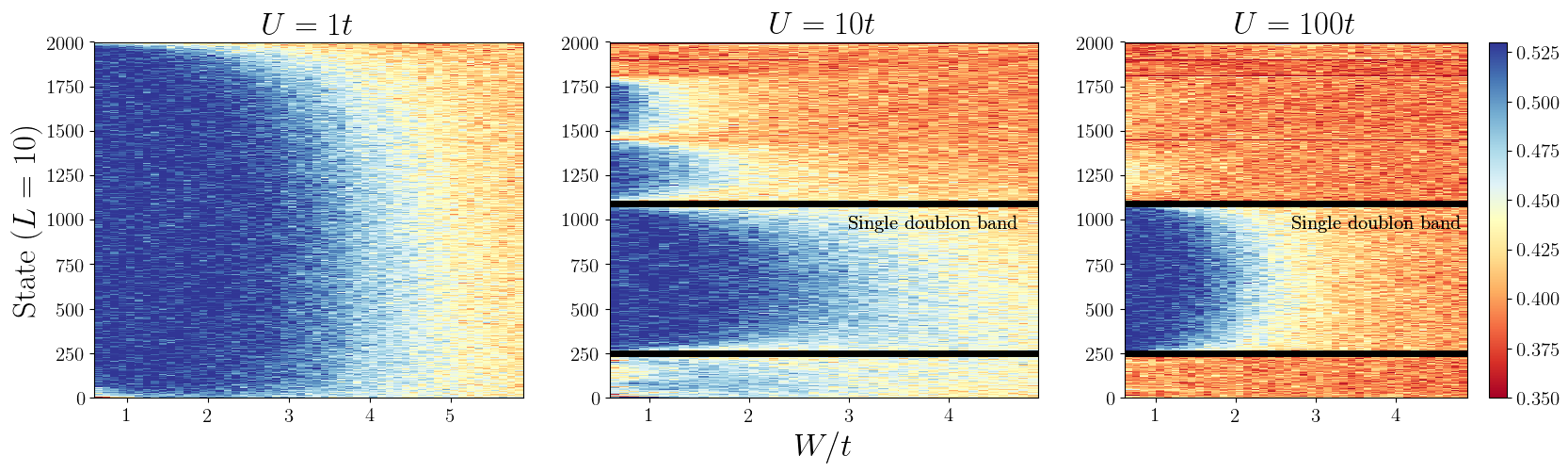}
\caption{Ergodic-to-MBL transition in a system of bosons on the lattice for increasing interaction strength. The plots show  the averaged gap-ratio $r$ as function of disorder strength $W/t$ and state label (see main text). The plots show clear emergence of Wigner-Dyson statistics (blue regions) versus Poisson statistics (yellow regions) when many-body states host a doublon. For $U=10t$ and $100t$ the many-body bands with one doublons are delimited by black horizontal lines. Numerics were performed on the Bose-Hubbard model~\eqref{eq:bh} with $L=10$ sites and $N=5$ bosons. Disorder averages were performed over 400 disorder realizations.}\label{fig:bosons}
\end{figure}

Figure~\ref{fig:bosons} is the analog of Fig.~1 in the main text, but derived for Eq.~\eqref{eq:bh}. As for Fig.~\ref{fig:finite}, also in this case, it is practical to exchange the axes and consider the state position in the many-body spectrum instead of its energy. This choice allows to better observe the WD-P transition in sectors with different numbers of doublons.   The phase diagram is totally analogous to the one derived for fermions in the main text. The main difference resides in the fact that more than one doublon and even triplons can be formed in the Bose-Hubbard model, being all bosonic particles indistinguishable. We remind here that, in our study concerning the Fermi-Hubbard model, we considered only a single spin-down fermion, thus allowing for the formation of a single doublon at most. For instance, the black lines drawn in Fig.~\ref{fig:bosons} for $U=10t$ and $100t$ sit exactly at $\binom{L}{N}$ and $\binom{L}{N}+L\cdot\binom{L-1}{N-2}$. Thus,  as in the fermionic case, it is possible to isolate the section of the Hilbert space with a single doublon and observe that its presence leads to Wigner-Dyson statistics also in bosonic systems. The similarity between the fermionic and bosonic case shows the generality of the mechanism discussed in the main text, that is the importance of doublon formation for the activation of thermalization in localized system.


\section{Details of the derivation of the effective single doublon Hamiltonian and analysis of resonances}
We provide here details concerning the description of a system with a doublon and the derivation of Eq.~(5) in the main text.
To describe a system with a doublon, we introduce the composite doublon annihilation operator $d_j=c_{j,\uparrow}c_{j,\downarrow}$ and then recast the onsite term in Eq.~(1) in the main text as $\mathcal H_{\rm O}=\sum_{j}\varepsilon_{j}n_{j,\uparrow}h_{j,\downarrow}+2\sum_j\varepsilon_jd^\dagger_j d_j$~\footnote{In this expression, we implicitly shift by $U$ the many-body spectrum.}, in which  $h_{j,\downarrow}=1-n_{j,\downarrow}$. 
We consider the case of large interaction  $U\gg N_\uparrow t$. This condition guarantees that many-body bands with and without doublon do not overlap in energy, see Fig.~1c in the main text.  We focus on the situation in which a doublon is formed and simplify Eq.~(1). We first separate the hopping term in Eq.~(1) in two parts $\mathcal H_{\rm T}=t(\mathcal D+\mathcal V)$, in which the operator  $\mathcal D(\mathcal V)=\sum_j[\mathcal D_j(\mathcal V_j)+\mbox{h.c.}]$ preserves(changes) the number of doublons~\cite{macdonald_$fractu$_1988,hofmann_doublon_2012,wurtz_variational_2020}
\begin{align}
\label{eq:t0} \mathcal D_j&=h_{j,\bar \sigma}c^\dagger_{j,\sigma}c_{j+1,\sigma}h_{j+1,\bar \sigma}+n_{j,\bar \sigma}c^\dagger_{j,\sigma}c_{j+1,\sigma}n_{j+1,\bar \sigma}\,, &
\mathcal V_j&=h_{j,\bar \sigma}c^\dagger_{j,\sigma}c_{j+1,\sigma}n_{j+1,\bar \sigma}+n_{j,\bar \sigma}c^\dagger_{j,\sigma}c_{j+1,\sigma}h_{j+1,\bar \sigma}\,,
\end{align}
in which   $n_{j,\bar \sigma}=c^\dagger_{j,-\sigma}c_{j,-\sigma}$ and $h_{j,\bar \sigma}=1-n_{j,\bar \sigma}$ are projectors controlling whether fermion hoppings change or preserve the doublon number.  The effective doublon-conserving Hamiltonian is obtained by eliminating $\mathcal V$ to leading order, which is achieved via a unitary Schrieffer-Wolff  transformation $\mathcal H'=e^{i S}\mathcal He^{-iS}$~\cite{schrieffer_relation_1966}  
\begin{align}\label{eq:sw}
\mathcal H'&=\mathcal H_{\rm O}+\mathcal D+\frac i2 \Big[S,\mathcal V\Big]+\mathcal O\big(S^2\big)\,,
\end{align}
where  $S$ is chosen to satisfy $ \mathcal V+i\Big[S,\mathcal H_{\rm O}+\mathcal D\Big]=0$.  In the particular case $N_\downarrow=1$, $\mathcal V$ only connects states $|\Psi_{1/0}\rangle$ with one/zero doublons. Such states are  of the form $|\phi_1\rangle=|\ldots ,\uparrow\downarrow,0,\ldots\rangle$ and $|\phi_0\rangle=|\ldots,\uparrow,\downarrow,\ldots \rangle$, in which we use the notation $|\uparrow\downarrow\rangle$ for the doublon at site $j$ and $|0\rangle$ for no particle at site $j+1$. They differ in energy by $U+\varepsilon_j-\varepsilon_{j+1}\sim U$, in the large $U$ limit. By evaluating the matrix element between such states for $\mathcal V$, we find
\begin{equation}
\langle \phi_0|\mathcal V|\phi_1\rangle+iU\langle \phi_0|S|\phi_1\rangle+\langle \phi_0|\Big[S,\mathcal D\Big]|\phi_1\rangle=0\,.
\end{equation}
As  $\langle \phi_0|S|\phi_1\rangle$ is of order $\mathcal O(t/U)$, the last term in the above expression is of order $\mathcal O(t/U)^2$, and we can neglect it in a first approximation. Accordingly, also the commutator $[S,\mathcal V]=\mathcal O(t^2/U)$, therefore  $\mathcal H'=\mathcal H_{\rm O}+\mathcal D+\mathcal O(t^2/U)$. We can also write $\mathcal H_{\rm O}=\sum_{j}\varepsilon_{j}n_{j,\uparrow}h_{j,\downarrow}+2\sum_j\varepsilon_jd^\dagger_j d_j$,   in which we  introduce the composite doublon annihilation operator $d_j=c_{j,\uparrow}c_{j,\downarrow}$. Further, the doublon-conserving operator $\mathcal D$ in Eq.~\eqref{eq:t0} can be considerably simplified in the single doublon sector: we can neglect operators of the form $h_{j,\uparrow}c^\dagger_{j\downarrow}c_{j+1,\downarrow}h_{j+1,\uparrow}$ (no site occupied exclusively by a spin-down) and $n_{j,\downarrow}c^\dagger_{j,\uparrow}c_{j+1,\uparrow}n_{j+1,\downarrow}$ . One thus obtains the residual hopping
\begin{equation}\label{eq:d1}
\begin{split}
\mathcal D=t \sum_j\big[ h_{j,\downarrow}c^\dagger_{j,\uparrow}c_{j+1,\uparrow}h_{j+1,\downarrow}+d^\dagger_jc_{j,\uparrow}c^\dagger_{j+1,\uparrow}d_{j+1}+\mbox{h.c.}\big]\,,
\end{split}
\end{equation}
where $h_{j,\downarrow}$ acts as a projector preserving the doublon number during singlon hopping. Expression~\eqref{eq:d1} clearly shows that, for $U\rightarrow\infty$, doublons can still hop with an amplitude of order $t$, only if singlons are present  nearby. Such correlated hopping  competes with localization, favoring ergodicity for weak enough disorder, as  sketched in Fig.~4 in the main text and as we are going to illustrate in detail below.

\subsection{Analysis of resonances } 

The correlated hopping~\eqref{eq:d1} efficiently induces resonances between initially fully localized configurations, below a critical disorder strength $W_{\rm C}$. To analyze this effect for moderate disorder, we make two simplifying assumptions: ({\it i})  doublons and singlons are independent particles that correspond to $d$ and $c_\uparrow$ annihilation operators acting on separate Fock spaces and ({\it ii}) we identify $h_{j,\downarrow}=\mathbb I$ in Eq.~\eqref{eq:d1}. The first approximation applies in the $U\rightarrow \infty$ limit. The second one neglects all local change of potential seen by singlons, following doublon displacements. Such processes can be neglected to leading order in $t$, in which the doublon acts as an infinite barrier between two AL systems composed  of free singlons.   These assumptions lead to an effective Hamiltonian describing the single doublon sector
\begin{equation}\label{eq:heff}
\mathcal H_{\rm eff}=\mathcal H_\uparrow+\sum_j\Big[2\varepsilon_j d^\dagger_jd_j+t d^\dagger_jc_{j,\uparrow}c^\dagger_{j+1,\uparrow}d_{j+1}+\mbox{h.c.}\Big]\,,
\end{equation}
in which $\mathcal H_\uparrow=\sum_j\varepsilon_jn_{j,\uparrow}+t\sum_j[c^\dagger_{j,\uparrow}c_{j+1,\uparrow}+\mbox{h.c.}]$ is  a single-particle nearest-neighbor hopping Hamiltonian  in which onsite disorder leads to AL.  It is thus convenient to switch to the basis of single-particle localized orbitals denoted by operators $a_l$, such that $c_{j,\uparrow}=\sum_l\psi_l(j)a_l$, where $\psi_l(j)\sim e^{-|j-l|/\xi}/\sqrt{\xi}$ are eigenfunctions localized around site $l$, with a typical exponential decay. The localization length $\xi$ scales as $t^2/W^2$ 
 for $W\ll t$~\cite{thouless_relation_1972,*czycholl_conductivity_1981,*kappus_anomaly_1981}. The effective Hamiltonian~\eqref{eq:heff} becomes
\begin{equation}\label{eq:heffal}
\mathcal H_{\rm eff}=\sum_l\mathcal E_la^\dagger_la_l+2\sum_j\varepsilon_jd^\dagger_jd_j
+t\sum_{j,l,m}\psi_l(j)\psi^*_m(j+1)\Big[d^\dagger_ja_la^\dagger_md_{j+1}+\mbox{h.c.}\Big]\,,
\end{equation}
in which $\{\mathcal E_l\}$ are the eigenenergies corresponding to AL states with  Poissonian level-spacing distribution.

We inspect the matrix elements between typical states $|\psi\rangle$ and $|\psi'\rangle$, sketched in Fig.~4b in the main text, with a doublon and $N_\uparrow$  localized orbitals randomly occupied  by spin-up fermions. To leading order, the correlated hopping term in Eq.~\eqref{eq:heffal} couples $|\psi\rangle$ to any state $|\psi'\rangle$ in which the doublon is displaced by one site and a particle-hole pair is created. For typical initial(final) states, the probability of occupied or empty localized single-particle orbitals  is $\rho_\uparrow$ and $1-\rho_\uparrow$ respectively, thus we can estimate the matrix element between these two states as 
\begin{equation}\label{eq:matel}
\langle \psi'|\mathcal H_{\rm eff}|\psi\rangle\simeq-t\psi_l(j)\psi_m(j+1)\rho_\uparrow(1-\rho_\uparrow)\,.
\end{equation}
Notice that this matrix element is suppressed, signaling doublon localization, both in the $\rho_\uparrow\rightarrow0$ and $\rho_\uparrow\rightarrow1$ limits. The nature of localization in such two limits is slightly different. For $\rho_\uparrow\rightarrow0$, localization is caused by the impossibility for doublons to hop  in the absence of singlons nearby. For $\rho_\uparrow\rightarrow1$, Eq.~\eqref{eq:heff} maps onto a tight-binding model for a single spin-down  with random on-site potential and hopping, subject to AL.  Notice that the matrix element~\eqref{eq:matel} is maximum at half-filling $\rho_\uparrow\approx1/2$. 

We compare now the matrix element~\eqref{eq:matel} with the relevant level spacing. To leading order, the number of states connected to $|\psi\rangle$ is roughly given by the number of ways  particle-hole excitations can be arranged in a localization volume $\xi$, thus $\propto\xi^2$. As a consequence, the average level spacing is given by $\delta\varepsilon=\delta E/\xi^2$, in which  $\delta E\simeq3W$ is the typical energy difference between the states $|\psi\rangle$ and $|\psi'\rangle$. By assuming that $\psi_l(j)$ are  oscillating functions
of amplitude $1/\sqrt\xi$ within a localization volume, the resonance condition (5) in the main text is derived. We report it here for readability
\begin{equation}\label{eq:transition}
3W\leq t\rho_\uparrow(1-\rho_\uparrow)\sqrt{\xi(W)}\,.
\end{equation}
This result shows how, surprisingly, a single doublon efficiently induces many-body resonances in a localized system below a critical disorder strength $W_{\rm C}$ -- defined by setting the equality in Eq.~\eqref{eq:transition} -- which strongly depends on the singlon density $\rho_\uparrow$. Notice that the  transition to the MBL phase for strong disorder is controlled by a single parameter $W/t$, the only one left in the $U\rightarrow\infty$ limit.

\bibliographystyle{apsrev4-1}
\bibliography{biblio}